# Data-Driven Investment Decision-Making: Applying Moore's Law and S-Curves to Business Strategies

Christopher L. Benson[1*], Christopher L. Magee[1,2]

[1]International Design Center, Massachusetts Institute of Technology
77 Massachusetts Avenue, Cambridge, MA 02139, USA

[2]MIT Institute for Data, Systems, and Society
77 Massachusetts Avenue, Cambridge, MA 02139, USA

*Corresponding Author
E-mail addresses: (CLB) cbenson@mit.edu, (CLM) cmagee@mit.edu

**Abstract**

This paper introduces a method for linking technological improvement rates (i.e. Moore's Law) and technology adoption curves (i.e. S-Curves). There has been considerable research surrounding Moore's Law and the generalized versions applied to the time dependence of performance for other technologies. The prior work has culminated with methodology for quantitative estimation of technological improvement rates for nearly any technology. This paper examines the implications of such regular time dependence for performance upon the timing of key events in the technological adoption process. We propose a simple crossover point in performance which is based upon the technological improvement rates and current level differences for target and replacement technologies. The timing for the cross-over is hypothesized as corresponding to the first 'knee' in the technology adoption 'S-curve' and signals when the market for a given technology will start to be rewarding for innovators. This is also when potential entrants are likely to intensely experiment with product-market fit and when the competition to achieve a dominant design begins. This conceptual framework is then back-tested by examining two technological changes brought about by the internet, namely music and video transmission. The uncertainty analysis around the cases highlight opportunities for organizations to reduce future technological uncertainty. Overall, the results from the case studies support the reliability and utility of the conceptual framework in strategic business decision-making with the caveat that while technical uncertainty is reduced, it is not eliminated.

Moore's Law, S-Curves, Technological Change, Investment, Competitive Positioning

# 1. Introduction

As technology continues to accelerate and becomes more influential in society and business, the need to better forecast the expected benefits and costs of these technological changes and their impact on business will increase. There has been extensive research focusing on what causes technological change (Jiang et al, 2010). The process by which technology diffuses has also been studied at length (Adner and Kapoor, 2016). There have been others who have focused specifically on how organizations can respond to technological change (Roy and Sarkar, 2014; Aggarwal et al, 2016). This leaves the question of when will technological change start affecting technological adoption, when will the dominant design be decided through trial and error, and when will firms need to execute their strategies for responding to rapid technological change.

# 2. Literature and Conceptual framing

Some theories of technological change and its impact on businesses focus on potential differences in the rates of improvements in technologies Dosi (1982), Foster (1986). The nature and causes of these differences in rates have been described in distinct frameworks such as exhaustion of possibilities for improvement in older technologies along with acceleration of improvement in new technologies (Utterback, 1974), or the theories of over-performance (Bower and Christensen, 1995), architectural change (Henderson and Clark, 1990) and other extensions of these ideas (Adner, 2002; Gans, 2006). In this paper, we do not focus on explanations (exhaustion, recombination, etc.) but instead on how to apply what has been found empirically about rates of performance change to corporate strategy concerning technological change. In doing so, we will be particularly aiming at questions regarding timing of key aspects of technological change. Therefore, we will consider the timing of adoption of products/services and the timing of performance improvements for technologies that are embedded in the products and services.

## 2.1 S-Curves

Nearly all modern theories of technological adoption/diffusion incorporate the 'S-curve' of adoption obtained when the number of users are plotted against time (Geroski, 2000; Hall, 2004; Hall et al, 2005) and apply different theories as to what causes this logistic function.
Such theories include economic based theories such as Griliches (1957) or Mansfield (1961), Roger's Theory on innovation diffusion (1962), The Concerns Based Adoption Model (CBAM), the Technology Acceptance Model and the Universal Technology Adoption and Use Theory (UTAUT). For more extensive coverage of these, we refer to Straub's (2009) excellent review of the main technology adoption-diffusion theories. The S-curve of adoption is one basic timing framework we employ in this paper. In particular, the beginning of the S-Curve is of interest due to that time being an area of intense competition between technologies and market participants looking to define the next dominant design (Adner and Kapoor, 2016; Chen et al, 2017).

**2.2 Technological performance change**

Regarding performance changes, we utilize the approach initiated by Gordon Moore more than 50 years ago for Integrated Circuits that performance improves exponentially with time. The generalization of his finding to all technological domains is referred to as the Generalized Moore's Law or technological improvement curve: this is the second timing framework that we use in this research (Moore, 1965; Sahal, 1979; Koh and Magee, 2006; 2008; Nordhaus (1997, 2007), Magee et al., 2016). Of importance is the fact that different technological domains exhibit vastly different *rates* of improvement in performance that are roughly time independent: Magee et al demonstrate a range between 3.1% and 65% per year (Magee, 2016). In the past, finding such technological improvement rates has been very resource intensive and for some domains nearly impossible (Benson, 2014). Recent work has developed the capability to estimate the key GML parameter for *arbitrary* domains from patents relatively quickly and easily, making the use of the GML more feasible in strategic business decision making (Benson and Magee (2015, 2016), Triulzi et al, 2018).

**2.3 Linking S-Curves and performance change**

A common attribute of many of the theories of technological diffusion is the close interdependence between technological diffusion (S-Curve) and the rate of technological change, as described by Bettis and Hitt (1995) and Farzin et al (1998).

*'Both the rate of technological change and the speed of technological diffusion have increased significantly in recent years. These two changes reinforce each other and their effects can not be easily separated. '(Bettis and Hitt, 1995, The New Competitive Landscape, Strategic Management Journal)*

*'The importance of technological uncertainties become more evident once it is noted that the firm's decision about how soon to adopt innovations depends on how fast and by how much technology will advance over time' (Farzin Et Al, 1998, Optimal Timing of Technology Adoption)*

Indeed, the linking of diffusion and technological improvement now has empirical support, see Woo and Magee (2017). Furthermore, it is likely that the sensitivity to uncertainty in the rate of change of technological performance and its impact on the timing of technological adoption/diffusion makes difficult the evolution of business strategy recommended by Bettis and Hitt (1995):

*'Executives in technology-intensive firms and in firms that extensively use technology must develop a better understanding of the relationship of strategy to technological change and achieve a close integration of the two.' (Bettis and Hitt, 1995, The New Competitive Landscape, Strategic Management Journal)*

This paper builds upon recent methods that have been developed to decrease the uncertainty in the rate of technological change to aid in business decision making surrounding strategic investments in new technological or market areas that are highly reliant on a new technical capability. In this, we aim to help move towards the vision stated above by Bettis and Hitt.

**2.4 Hypothesis for use of performance improvement in business strategy**

As noted above, the key timing analytical framework that we will use is the adoption S-curve. We will focus on the early demand acceleration which is the first profit opportunity and a time of maximum competition to establish dominant designs (Abernathy and Utterback, 1978). Given the capability that now exists to quickly and inexpensively estimate the key generalized Moore's Law parameter for *arbitrary* domains from patents (Benson and Magee, 2015, 2016; Triulzi et al, 2018), we will show it is possible to link the two timing analytical frameworks for any desired domain. Our approach for doing this is given in the following steps:

1. Determine a technological domain that is a possible replacement technology (Replacement technology domain);

2. Determine the technical performance improvement curve of the replacement technology domain (by measurement or estimation)
3. Find the technology that currently performs the function (Target technology domain)
4. Determine the technical performance improvement curve of the Target Domain (by measurement or estimation)
5. Determine the feasibility-time-range that the performance in the Domain of Interest is likely to be approximately equal to the performance of the Target Domain
6. This feasibility-time-range will correspond to the first 'knee' in the technology adoption/diffusion curve.

In this paper, we seek to determine if this model is useful through evaluation of the hypothesis stated in step 6 of the process.

H1: The early knee of the technology adoption-diffusion S-curve occurs when the performance of an improving new technology reaches a level adequate to satisfy some customers –this is often equivalent to the performance level of the technology being replaced. This hypothesis is shown graphically in figure 1. In the top part of the figure, schematic performance curves are shown for the target and replacement technology and in the bottom part, a schematic adoption curve for the replacement technology is shown. The working hypothesis is that the S curve acceleration corresponds in time to the cross-over in the performance curves. The empirical examination of this hypothesis is the main purpose of this paper. Section 3 describes the data collection for both cases and then describes the uncertainty analysis surrounding the key parameters used in testing the hypothesis. Section 4 gives the results including the uncertainty which overall indicates that the cases support the hypothesis. Section 5 discusses the implications of the finding and opportunities for reducing uncertainty and section 6 concludes.

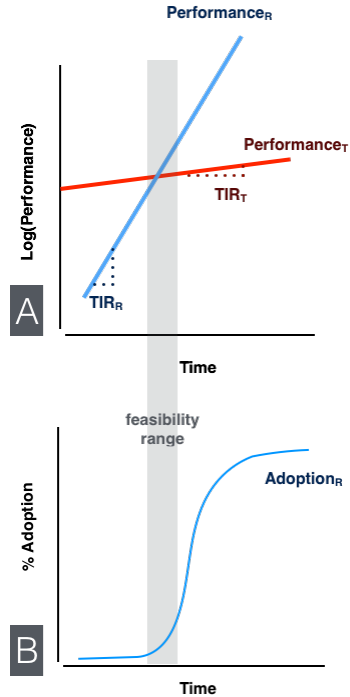

**Figure 1: Graphical depiction of research hypothesis**

## 3. Data

To test the hypothesis stated in Section 2 and shown graphically in figure 1, data is analyzed concerning the technological improvement and adoption over time for two domains. The overall intent is to test whether the crossover in performance is a good signal for the acceleration of the adoption. This section will describe the data sources used to construct the performance improvement and adoption curves and the required data reduction to test the hypothesis. The section finishes with describing the sources of uncertainty that that are present in the data that potentially limit the generalization of the hypothesis.

We follow the definition previously given definition (Benson et al, 2018) that technological domains consist of artifacts having a specific 'useful purpose' (i.e. What utility does the domain provide for society and/or customers) and specific 'underlying scientific phenomena' used by the invention (i.e. How does the invention accomplish that utility?). Table 1 shows the 4 domains analyzed in this paper along with their purpose and underlying phenomena.

**Table 1: Replacement and target domains**

| Domain | Useful Purpose | Scientific Phenomena |
|---|---|---|
| Internet Audio Distribution (Replacement) | Distributing audio information | Wired information transmission |
| CD mail-order distribution (Target) | Distributing stored audio information | Physical matter movement |
| Internet Video Distribution (Replacement) | Distributing video information | Wired information transmission |
| DVD mail-order distribution (Target) | Distributing video information | Physical matter movement |

The domains are grouped as replacement and target domains by their common useful purposes so their performance can be compared (i.e. Internet Audio vs CD Mailing, Internet Video vs DVD Mailing) to find the 'feasibility range' and thus correlate with the first S-curve "knee".

**3.1 Calculating the technological improvement curve of the replacement technology**

This section describes the data and methodology for constructing the technological improvement curve for the replacement technology, which is represented by the 'Performance$_R$' curve in figure 1A. To determine the technological improvement curve, the first parameter to consider is the measure of technological performance, which is represented by the vertical axis in Figure 1A. For this analysis, the measure of performance is the amount of information (whether audio or video) that can be transmitted per dollar. This performance parameter can be quantified for both audio and video media types. The cost of transmitting information over the internet is calculated using Equation 1.

**Equation 1: Calculating Technical Performance of Internet Media Distribution**

$$InternetDistributionCost(t) = \frac{\#SecondsInMonth}{InternetSpeedCost(t)} * \frac{CompressionRatio_{Media}(t)}{FileSize_{Media,Uncompressed}}$$

Where *InternetDistributionCost(t)* is the amount of information that can be transmitted per dollar *(#MediaUnits/$)* and is measure of technical performance for the replacement domains.

*InternetSpeedCost(t)* is the inflation-adjusted cost per megabit of internet speed per month (i.e. if one's monthly internet speed is 5 Megabits[1] per second and their monthly bill is $25, their internet speed cost is 5 $/Mbps). Values of this parameter are given in Appendix Table A.1 from 1983 until 2010 along with the referenced source where the values were found (Coffman and Odlyzko, 1998; Odlyzko, 1998; Norton, 2010).

*#SecondsInMonth* is the number of seconds in a 30-day month (2,592,000).

*CompressionRatio(t)* is the compression ratio which differs among media types. The values of this parameter for audio and visual media are shown in Table A7 from the data found in Hilbert and Lopez (2011).

*FileSize* is the reference uncompressed file size for the type of media used. The baseline[2] audio file size is based on a 60 minute audio album and can be calculated using Equation 2 and the reference video file is one five-minute standard definition video clip whose size can be calculated using equation 3.

**Equation 2: Reference Audio File Size**

$$AudioFileSize = BitRate * LengthOfAudioFile$$

The bitrate (quality) from Hilbert and Lopez (2011) is used for an uncompressed bit rate of 633.6 kilobits per second, which gives an uncompressed 60 minute album a size of 2,280.96 megabits using equation 2.

**Equation 3: Reference Video File Size**

$$VideoFileSize = (PixelHeight * PixelWidth * BitsPerPixel * FramesPerSec + AudioBitRate) * LenVideo$$

Standard definition video has a *PixelHeight* of 480 pixels, a *PixelWidth* of 640 pixels, 24 *BitsPerPixel*, 30 *FramesPerSec* and an *AudioBitRate* of 633.6 kilobits per second, which gives the uncompressed *VideoFileSize* of 66.545 gigabits.

Using equations 1, 2 and 3 along with Tables A.1 and A.2, the technological performance curve for the replacement domains is calculated over time and is presented in the Results section.

---

[1] Please note that it is important in these types of analyses to pay close attention to the difference between a bit and a byte and to remember that 1 byte = 8 bits

[2] Other reference file sizes will be considered in the uncertainty analysis and reported in the Results Section.

## 3.2 Calculating the technological improvement curve of the target technology

The next step in the analysis is to calculate the technological improvement curve for the target domain, which is represented in Figure 1(A) by the curve labeled *Performance$_T$*. The target domains listed in Table 1 are evaluated using the same performance metric as the replacement domains: amount of media distributed per dollar, however the method of distribution is through the mail rather than through the internet. The cost of transmitting media through the mail is calculated using equation 4.

**Equation 4: Calculating Technical Performance of Mail Media Distribution**

$$MailDistributionCost(t) = \frac{1}{PostageCost_{FirstOuce}(t) + (Weight_{PhysicalStorage} - 1) * (PostageCost_{AdditionalOunce}(t))}$$

*MailDistributionCost* is the technical performance measure for the target domain and is measured by the number of media units that can be distributed per inflation-adjusted dollar.

*PostageCost$_{FirstOunce}$* is the amount in inflation adjusted dollars for the cost of mailing this first ounce of a letter and *PostageCost$_{AdditionalOunce}$* is the amount for additional ounces. The source and the values for this parameter can be found in Appendix Table A.3 (US Post Office Historian, 2018; Wikipedia, 2018f).

*Weight$_{PhysicalStorage}$* is the weight in ounces of the physical media rounded up to the nearest ounce, which in this case is either CDs or DVDs and both weigh <1 ounce in a sleeve.

Using Equation 4 and Table A.3, the technological performance curve for both target domains were calculated and are reported in the results section.

## 3.3 Calculating the adoption curve of the replacement technology

We determine adoption rates of specific technologies using the % adoption of that technology compared with its competition (thus audio is compared to other audio modes), this is represented by the *Adoption$_R$* curve in Figure 1B.

We calculate the total adoption of audio and video by the percent of number of minutes of media that are distributed by each method. This normalizes for the length of song, number of songs on an album, and for compression rates used on digital products and is consistent with Hilbert and Lopez's measurements.

Therefore, the information adoption rate of each technology/media was calculated using equation 5:

**Equation 5: Calculating the Adoption Curve for Internet Media**

$$Adoption_{Internet} = \frac{\#Minutes_{Internet,Media}}{\sum \#Minutes_{OtherDomains,Media}}$$

### 3.3.1 Usage of internet media

Equation 6 is used to calculate the number of minutes of media transferred over the internet, which is the numerator in equation 5 above.

**Equation 6: Calculating Usage of Media Distributed over the Internet**

$$\#Minutes_{Internet,Media}(t) = \frac{TotIntUse(t) * \%IntUse_{Media}(t)}{\left(\frac{FileSize_{Media,1Min,Uncompressed}}{CompressionRatio_{Media}(t)}\right)}$$

*TotIntUse* is the total usage of the internet, measured in Gigabytes per year and is given in Appendix Table A.4 along with the source of this data (Sumits, 2015).

*%IntUse* is the percent of total internet use by media type, which is derived from the usage data from Hilbert and Lopez (2011) and is found in Appendix Table A.5.

*FileSize* is the uncompressed file size and was calculated using Equation 2 for Audio and Equation 3 for video using 1-minute as the media length.

*CompressionRatio* is the same as the compression ratio in Equation 2 and values/sources are found in Appendix A.1.

### 3.3.2 Usage of physical media

This section describes the calculation of the usage of alternate technology domains for audio and video distribution, the sum of which is represented by the numerator in equation 5.
For music, the alternate domains considered are:
- Vinyl (records)
- Cassettes
- CDs

And for video:
- VHS
- DVDs

For the analog media types including vinyl, cassettes, and VHS the usage metric can be calculated using equation 7.

**Equation 7: Usage calculation for analog physical media**

$$\#Minutes_{Domain,Media} = YearlySales_{Domain} * MinutesPerUnit_{Domain}$$

Where *YearlySales* is the number of total sales of a specific type of media (i.e. audio cassettes) and is shown in Appendix Table A.6 along with the sources of this data (Hilber and Lopez, 2011).

*MinutesPerUnit* is the number of minutes on each data unit, which is shown in Appendix Table A.7.

For the digital storage media including CDs and DVDs, the usage metric was calculated using equation 8.

**Equation 8: Usage calculation for digital physical media**

$$\#Minutes_{Domain,Media} = \frac{YearlySales_{domain} * UnitDataStorage_{domain} * CompressionRatio_{Media}(t)}{FileSize_{Media,1Min,Uncompressed}}$$

Where *UnitDataStorage* is the amount of data that each unit of the digital physical media can store and is shown in Appendix Table A.8.

*CompressionRatio* and *FileSize* are the same as in Equation 6.

**3.4 Sources of uncertainty**

In this section, we identify all sources of uncertainty in our analyses. In general, there is uncertainty in the 'overlap' of the technological performance curves and uncertainty (even after the fact) of the time range of the "acceleration knee" in the adoption curve. These sources of uncertainty are shown in Figure 2 and are described in numerical order in this section. The calculated uncertainties will be reported in the Results section.

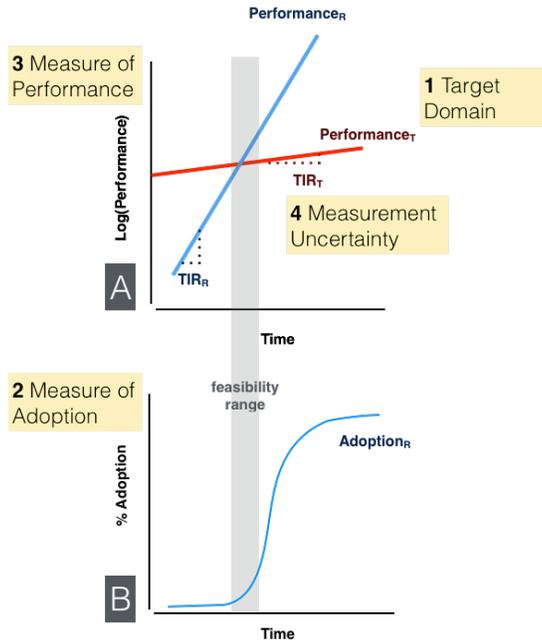

**Figure 2: Sources of uncertainty**

### 3.4.1 Target domain definition

Distributing media through the internet represents the replacement domains of interest for this paper, and two target domains were sending digital physical media through the mail. The selection of alternate target domains such as sending analog media (i.e. VHS or cassettes) through the mail or driving to the store to pick up any sort of physical media (i.e. Blockbuster) would affect the analysis and thus our choice of target domains introduces some uncertainty in the target domain performance and therefore introduces uncertainty about when the replacement domain performance becomes equivalent (crossover).

### 3.4.2 Measure of adoption

The metric used to define usage of a technology can also contribute to uncertainty. For this paper we used compression adjusted data that is number of minutes of video or audio. Alternative measures of adoption include the percent of uncompressed 'raw' data per domain, or the number of units transmitted (i.e. a song or a movie) and introduce uncertainty concerning the timing of the measured knee of the adoption curve.

### 3.4.3 Measure of performance

Selection of the technological performance reference metric also introduced uncertainty. While the cost to transmit an album and a standard-definition video clip were chosen, other reference units could be a song or a full-length high-definition movie or many options in between. Additionally, we chose to focus intently on the distribution of the media, and not on the manufacture or final consumption. Other performance measures could include aspects such as the cost of manufacturing a CD or DVD, or the yearly cost of a computer, DVD player, modem, or cassette player.

### 3.4.4 Performance measurement uncertainty (i.e. the slope of the technological improvement curve)

Finally, just as in any measure, there will always be errors in the measurement of the data. Since we are looking back in time in the two test cases, we use the actual technological performance for each of the domains. The actual measures are noisy and thus there could be reason to use a regression line for the technical improvement rate (TIR) as is denoted by TIR in Figure 1. For future predictions, a forecast of the technological improvement rate would arise from a regression of patent parameters to yield a predicted TIR and we consider this interesting type of application in the Conclusion section.

## 4. Results

### 4.1 Feasibility range of internet audio transmission and internet video transmission

The test of the hypothesis is achieved by comparing the technological performance 'cross-over' point ("feasibility range") with the 'knee' in the technological adoption curve. This section will describe the results for distribution of audio and video over the internet compared with the baseline target domains that they replaced. Figure 3A shows the costs of transmitting audio over the internet (replacement domain) and the same cost for mailing the same audio (the target domain) and Figure 3B shows the rise in internet audio adoption in the same time period. These graphs look reasonably like the theoretical model depicted in Figure 1 with the 'cross-over' of the technological performance curves (Figure 3A) occurring in 1998, followed shortly thereafter by the sharp increase in internet audio adoption in 1999 (Figure 3B).

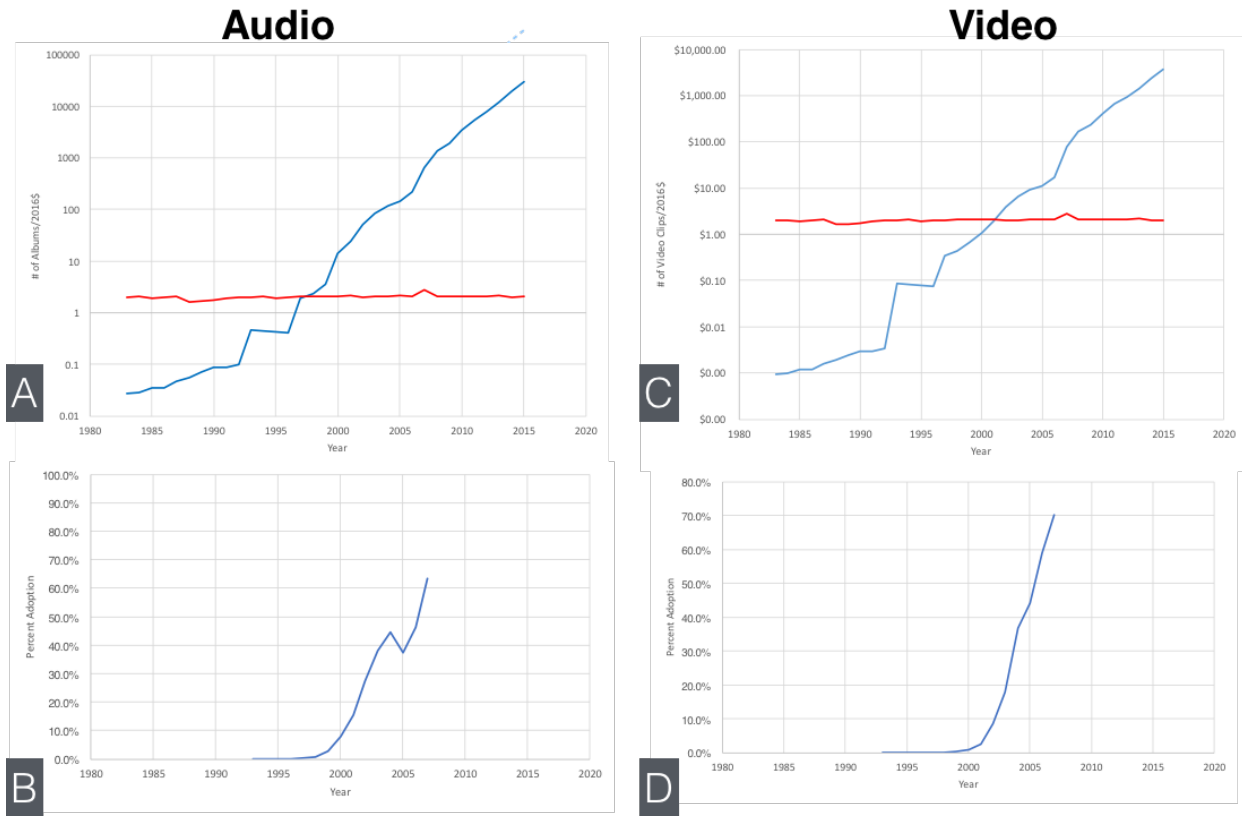

**Figure 3: Internet media performance and adoption comparison**

Figure 3C shows the costs of transmitting a video over the internet (replacement domain) and the same cost for mailing the same video (the target domain). The crossover is seen to occur in 2002. Figure 3D indicates that the acceleration knee also occurs in 2001.
Overall, Figure 3 offers support for the hypothesis given that the adoption increase is within a year or so of the performance cross-over. Before analyzing the robustness and uncertainty of such conclusions, it is useful to briefly summarize a qualitative analysis of relevant events that happened in this same period.

One important date to consider relative to the quantitative audio analysis is the founding of Napster in 1998 - a date very close to the crossover date seen in Figure 3A (Napster, 2018). A second date to note is the initial launch of iTunes in 2001-this was an update of a program developed and first released in 1999 by Cassady and Greene and purchased by Apple in 2000 (SoundJam MP, 2018). These dates are also close to expectations from Figure 3A.

The qualitative changes for the video case are more diverse and not easily linked to the quantitative analysis in Figures 3C and 3D. One factor in the complexity of comparison is that the dominant target domain changed during the time of the analysis. Although DVD's reached the US market in 1997 (DVD, 2018), peak DVD consumption did not occur until 2005 (Fellows, 2013). Similarly, Netflix was founded in 1997, successfully had an IPO in 2002 based upon a DVD mailing model, announced streaming video as late as 2007 and then soon (and unsuccessfully) tried to sell their DVD business in 2011(Netflix, 2018). Blockbuster's founding (1985) was well before the relevant video dates in Figures 3C and 3D but this is reasonable since their original business (rental in physical outlets) was a possible target for Internet video modes rather than a beneficiary. Consistent with this is the fact that they were slow to offer DVD mailing (2004) and their large physical retail burden slowly pushed them to a bankruptcy declaration in 2010 (Blockbuster LLC, 2018). Another noteworthy real world event was the successful launch of YouTube in 2005 (Youtube, 2018). Although none of these events is starkly contradictory to the quantitative analysis of the video transition, there are multiple possibilities among which some appear at least a little late (you tube and Netflix streaming) and others appear potentially early (Netflix founding) but most unsatisfactory is that the qualitative facts are at best only roughly and unclearly aligned with the quantitative analysis.

To summarize the hypothesis test to this point, the audio case is qualitatively and quantitatively in support of the hypothesis. On the other hand, the baseline crossover for Internet video is consistent with the acceleration of the adoption curve but the qualitative analysis is not well aligned with the quantitative analysis. Moreover, we will now see that uncertainties inherent in some of the parameters used to make this comparison are large enough to make the indicated support less robust than desired. We will now present these results before considering the implications of the findings for practical application including analyzing the relationship of the uncertainties that affect the parameters to strategic decision making by management.

**4.2 Uncertainty of Results**

The Data section identified several areas that introduce uncertainty into the testing of the hypothesis that was reported in the Results section. This subsection will provide quantification of the uncertainty analysis described earlier. In each topic affecting uncertainty, we will report results for the audio case and the video case in a single table

### 4.2.1 Uncertainty in selection of target domains

Selection of different target domains affect the 'cross-over' date as is shown in Table 2.

**Table 2: Uncertainty effects of different target domains (audio on the left, video on the right)**

|  | **Mail CD** | Drive | Mail Cassette | **Mail DVD** | Drive |
|---|---|---|---|---|---|
| Performance 'Cross-over' Year (w/ Internet) | **1998** | 1997 | 1997 | **2002** | 2001 |

On the left side of the table it is seen that driving to pick up a CD or mailing a cassette only changes the crossover date by 1 year relative to the baseline case (mailing a CD). Thus, in this case, the selection of the target domain contributes minimal uncertainty.

### 4.2.2 Uncertainty in selection of usage metric

Other measures for usage and their effect on the 'knee' in the usage curve are shown in Table 3 with the audio case on the left-hand side of the table and the video case on the right.

**Table 3: Uncertainty effects of different adoption measures**

|  | **Number of Minutes of Audio** | Raw Audio Data | Number of 'songs' (3-min) | **Number of Minutes of Video** | Raw Video Data | Number of 'Movies' |
|---|---|---|---|---|---|---|
| Usage Knee (above 1%) | **1999** | 1999 | 1999 | **2001** | 2002 | 2000 |
| Usage Knee (above 10%) | **2001** | 2001 | 2000 | **2003** | 2005 | 2002 |

Defining the 'knee' in the adoption as either 1% or 10% introduces a change in the feasibility range of only a year or two. Similarly, different metrics for measuring adoption, (particularly raw video) also only change the feasibility range by one or two years.

### 4.2.3 Uncertainty in different performance metrics

The effect of selecting different performance measurement units on the 'cross-over' date is shown in Table 4.

**Table 4: Uncertainty effects of different performance measures**

|  | **Album** | Song (3 min) | **Video Clip** | SD Movie (90 min) | HD Movie (90 min) |
|---|---|---|---|---|---|
| Uncompressed Size | **2,280MB** | 114MB | **66.5GB** | 1197GB | 3027GB |
| Performance 'Cross-over' Year (w/ CD/DVD-by-mail) | **1998** | 1992 | **2002** | 2007 | 2008 |

The difference between considering adoption of songs vs albums (our baseline) is significant as is video clips (our baseline) vs high definition movies. The results in the Table indicate that such changes in the necessary performance level changes the crossover date by as much as 6 years which appear to be sufficient to make the predictions from the hypothesis of much reduced value. As will be discussed in the Discussion and Conclusion section, the sensitivity of the measure of technological performance to the feasibility range presents significant opportunity for businesses to reduce future technological uncertainty by a greater understanding of this aspect of their technologies and products and how this relates to their understanding of their customers.

### 4.2.4 Uncertainty in performance measurement errors

Table 5 shows the impact on the 'cross-over' year when using the TIR exponential regression instead of the actual data points.

**Table 5: Uncertainty effects on performance measurement errors**

|  | **Empirical Data (Internet Audio)** | Exponential Regression (all data) (Internet Audio) | Exponential Regression (from 1995) (Internet Audio) | **Empirical Data (Internet Video)** | Exponential Regression (All Data) (Internet Video) | Exponential Regression (from 1995) (Internet Video) |
|---|---|---|---|---|---|---|
| Performance 'Cross-over' Year (Internet vs CD/DVD) | **1998** | 1996 | 2001 | **2002** | 2001 | 2002 |

As with the first two sources of uncertainty, there is an effect but its magnitude is not large enough to invalidate the support of the hypothesis seen in the baseline results.

### 4.2.5 Summary of uncertainty quantification

The summary of all the parameter uncertainty is graphically depicted in Figure 4 below with figures 4A showing the audio case and figure 4B showing the video case. Considering the audio case, the hypothesis is supported by the results, but are very sensitive to the choice of performance metrics. In the video case, the alignment of the crossover and acceleration knee is supported, but may show significant divergence if the target domain and performance metric vary to opposite extremes in their range. Overall, the results of the uncertainty analysis are consistent with the qualitative analysis summarized above.

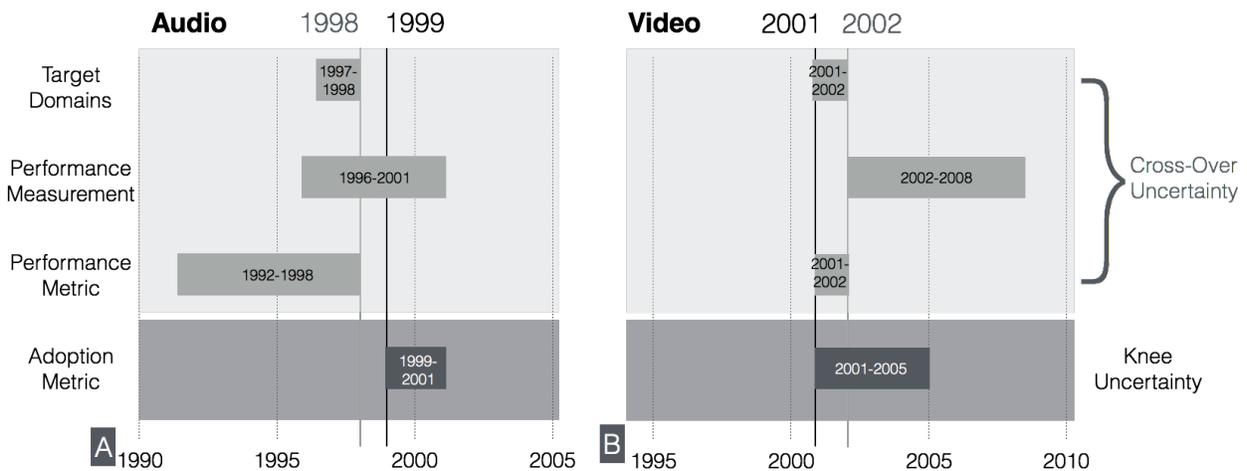

Figure 4: Summary of uncertainty of hypothesis tests

### 5. Discussion and Conclusion

Although the overall evidence supports the hypothesis that technological performance curves (i.e. Moore's Law) and technological adoption curves (i.e. S-Curves) are linked, the accuracy of predictions for the linkage is not found to be highly reliable. There is significant uncertainty in predicting the crossover for one of our two cases and the results are sensitive to many choices that have to be made by those making the forecast. In this section, we first discuss this finding emphasizing that the choices made in forecasting cross-over are homologous with analysis that needs to be done qualitatively to make a rational strategic decision about committing to a serious business strategy surrounding a significant technological change. We will then more broadly explore the implications of the demonstrated linkage outlining some potentially useful ways to use the framework in innovation management.

The variation in the predicted time for the crossover in performance is strongly dependent upon the baseline application assumed for the target domain. This is parallel to a critical decision that must be made by business leaders considering the timing for resource commitment. To make a proper selection of the target domain, it is important for innovation managers to determine who their main competitors are (even if they are internal).

Moreover, selecting an appropriate usage metric for use in prediction requires consideration of the position of the business in the value chain. In our examples, the appropriate usage metric is different for infrastructure investors such as ISP providers, Internet hardware suppliers (for example Cisco), or wireless service providers than it would be for a company proposing a service to download and sell music (Apple I store) and for one that wants to focus on small movies widely sourced (You Tube) or HD movies (BluRay). This analysis is also obviously necessary for business managers who must carefully determine their position in the value chain and how it affects the performance/cost they need to achieve to be successful against a target existing domain. For example – is the business selling data (i.e. ISP) or are they selling movies (i.e. Music Studios) or are they cost sensitive to the amount of time you spend watching a movie (i.e. Streaming Services). Based upon their interest they will have a different usage metric as their baseline.

The sensitivity to the measurement of technology performance improvement makes it critical for a business strategy team to understand exactly what their customers care about. It appears to us that the quantitative analysis of the predicted date for acceleration of adoption would provide a framework for the overall business analysis and would allow some quantification to be added to the analysis. For example, the inherent uncertainty in measurement of performance data provides incentive for the importance of business collecting quality data on the technical performance of their own and competing products and services. The added quantified prediction should be treated as valuable additional information and not as a substitute or replacement for deep qualitative analysis which can always take in additional factors not fully quantifiable. The additional information would come at little extra resource effort and thus is expected to be generally cost-effective (Young, 1989).

Given the alignment of the performance cross-over and the acceleration of adoption broadly seen in the current study, one may well wonder how broadly applicable the use of this technique can be. As we have just discussed, much of the analysis is work that should be done in making these kinds of decisions. The additional work is to project performance to future time periods. One approach to this is to use regression fits to Moore's Law as discussed above and this has been analyzed to show reasonable accuracy with modest time extrapolation (Farmer and Lafond, 2016). A major apparent limitation arises however since such performance measurement is time consuming and difficult so it has only been accomplished on a small minority of possible domains of interest. The limitation has recently been removed by use of a patent based technique which has evolved (Benson and Magee, 2015) to become quite reliable (Truilzi et al, 2018). Therefore, it is now possible to predict the technological improvement rate (or annual improvement %) for arbitrary domains. Thus, the following approach is applicable generally by businesses:

1. Through research, development and market analysis, determine a technical domain of interest, which is usually a technology that the business has access to or is considering investing in. (this will be the replacement technology)
2. Through competitive analysis, determine potential target technologies, even if they come internally.
3. Through business model and customer discovery, determine the appropriate measure for technical performance.
4. Determine the future technical improvement curve either through measurement (Farmer and Lafond, 2016) or if resources are constrained, through estimation (Benson and Magee, 2015, Triulzi et al 2018) for the replacement and target domains.
5. Include uncertainty when calculating the 'cross-over' to determine the 'feasibility time range' for when a given experimentation with a given technology toward a dominant design will begin in earnest.
6. Continue to update this forecast with new technology, competition, business model, and customer information, both quantitative and qualitative, and use it as a data point for when to enact business strategies surrounding significant technological change.

The emergence of reliable prediction of annual improvement percentages for arbitrary technologies makes wider use of them in strategic management a logical target for practically-oriented research and is the motivation for the research reported here. As we recommend such wider use it is also important that objective testing of the basic concept be undertaken and the two cases examined in this paper accomplish that task. These tests have given some important support to the conceptual framework but have also uncovered the kind of issues typically uncovered in initial applications-there are important definitional and analytical tasks that must be considered in using the approach and there is no guarantee that we have uncovered all of the issues. Thus, a major limitation of the current research is that its practical implications are not nearly fully known since we are at an early stage of its evolution in practice.

Triulzi G, Alstott J, Magee CL. 2018. Estimating Technology Performance Improvement Rates by Mining Patent Data. *SSRN Electronic Journal* : 1–54. Available at: https://www.ssrn.com/abstract=2987588.

US Post Office Historian. 2018. Rates for Domestic Letters Since 1863. Available at: https://about.usps.com/who-we-are/postal-history/domestic-letter-rates-since-1863.htm [26 April 2018].

Utterback JM. 1974. Innovation in industry and the diffusion of technology. *Science (New York, N.Y.)* **183**(4125): 620–6. Available at: http://www.ncbi.nlm.nih.gov/pubmed/17778831.

Young MA. 1989. Research notes and communications sources of competitive data for the management strategist. *Strategic Management Journal* **10**(3): 285–293. Available at: http://doi.wiley.com/10.1002/smj.4250100307.

DVD. 2018. *Wikipedia*. Available at: https://en.wikipedia.org/wiki/DVD [26 April 2018].

History of United States postage rates. 2018. *Wikipedia*. Available at: https://en.wikipedia.org/wiki/History_of_United_States_postage_rates [26 April 2018].

Netflix. 2018. *Wikipedia*. Available at: https://en.wikipedia.org/wiki/Netflix [26 April 2018].

Blockbuster LLC. 2018. *Wikipedia*. Available at: https://en.wikipedia.org/wiki/Blockbuster_LLC [26 April 2018].

YouTube. n.d. *Wikipedia*. Available at: https://en.wikipedia.org/wiki/YouTube [26 April 2018].

Napster. 2018. *Wikipedia*. Available at: https://en.wikipedia.org/wiki/Napster [26 April 2018].

SoundJam MP. 2018. *Wikipedia*. Available at: https://en.wikipedia.org/wiki/SoundJam_MP [26 April 2018].

Woo J, Magee CL. 2017. Exploring the relationship between technological improvement and innovation diffusion: An empirical test. *arXiv* : 1–31. Available at: http://arxiv.org/abs/1704.03597.
25

**Appendix**

**Table A.1 Average Cost for Internet Bandwidth Over Time**

| Year | $/Mbps/Monthly bill | 2016$/Mbps/Monthly bill | Source |
|---|---|---|---|
| 1983 | $16,666.67 | $42,184.50 | Coffman and Odlyzko, 1998 |
| 1984 | $16,000.00 | $39,196.17 | Coffman and Odlyzko, 1998 |
| 1985 | $13,666.67 | $32,034.64 | Coffman and Odlyzko, 1998 |
| 1986 | $14,333.33 | $32,400.87 | Coffman and Odlyzko, 1998 |
| 1987 | $11,000.00 | $24,403.59 | Coffman and Odlyzko, 1998 |
| 1988 | $9,666.67 | $20,662.89 | Coffman and Odlyzko, 1998 |
| 1989 | $7,833.33 | $16,051.31 | Coffman and Odlyzko, 1998 |
| 1990 | $6,666.67 | $13,002.48 | Coffman and Odlyzko, 1998 |
| 1991 | $7,166.67 | $13,222.43 | Coffman and Odlyzko, 1998 |
| 1992 | $6,500.00 | $11,487.78 | Coffman and Odlyzko, 1998 |
| 1993 | $5,333.33 | $9,142.12 | Coffman and Odlyzko, 1998 |
| 1994 | $5,666.67 | $9,422.72 | Coffman and Odlyzko, 1998 |
| 1995 | $6,166.67 | $9,991.58 | Coffman and Odlyzko, 1998 |
| 1996 | $6,500.00 | $10,233.20 | Coffman and Odlyzko, 1998 |
| 1997 | $1,466.00 | $2,239.83 | Odlyzko, 1998 |
| 1998 | $1,200.00 | $1,791.35 | Norton, 2010 |
| 1999 | $800.00 | $1,175.63 | Norton, 2010 |
| 2000 | $675.00 | $970.03 | Norton, 2010 |
| 2001 | $400.00 | $555.51 | Norton, 2010 |
| 2002 | $200.00 | $269.85 | Norton, 2010 |
| 2003 | $120.00 | $159.35 | Norton, 2010 |
| 2004 | $90.00 | $116.79 | Norton, 2010 |
| 2005 | $75.00 | $94.73 | Norton, 2010 |
| 2006 | $50.00 | $61.02 | Norton, 2010 |
| 2007 | $25.00 | $29.52 | Norton, 2010 |
| 2008 | $12.00 | $13.77 | Norton, 2010 |
| 2009 | $9.00 | $9.93 | Norton, 2010 |
| 2010 | $5.00 | $5.54 | Norton, 2010 |
| 2011 | $3.25 | $3.54 | Norton, 2010 |
| 2012 | $2.34 | $2.47 | Norton, 2010 |
| 2013 | $1.57 | $1.62 | Norton, 2010 |
| 2014 | $0.94 | $0.96 | Norton, 2010 |
| 2015 | $0.63 | $0.63 | Norton, 2010 |

**Table A.2 Compression Ratios for Different Technology Types**
*Derived from Hilbert and Lopez (2011)*

| Year | Text | Image | Audio | Video |
| --- | --- | --- | --- | --- |
| 1983 | 1 | 1 | 1.00 | 1 |
| 1984 | 1 | 1 | 1.00 | 1 |
| 1985 | 1 | 1 | 1.00 | 1 |
| 1986 | 2.2 | 1 | 1.00 | 1 |
| 1987 | 2.2 | 1 | 1.00 | 1 |
| 1988 | 2.2 | 1 | 1.00 | 1 |
| 1989 | 2.2 | 1 | 1.00 | 1 |
| 1990 | 2.2 | 1 | 1.00 | 1 |
| 1991 | 2.2 | 1 | 1.00 | 1 |
| 1992 | 2.2 | 1 | 1.00 | 1 |
| 1993 | 2.9 | 7 | 3.68 | 20 |
| 1994 | 2.9 | 7 | 3.68 | 20 |
| 1995 | 2.9 | 7 | 3.68 | 20 |
| 1996 | 2.9 | 7 | 3.68 | 20 |
| 1997 | 2.9 | 7 | 3.68 | 20 |
| 1998 | 2.9 | 7 | 3.68 | 20 |
| 1999 | 2.9 | 7 | 3.68 | 20 |
| 2000 | 4.6 | 14 | 12.00 | 27 |
| 2001 | 4.6 | 14 | 12.00 | 27 |
| 2002 | 4.6 | 14 | 12.00 | 27 |
| 2003 | 4.6 | 14 | 12.00 | 27 |
| 2004 | 4.6 | 14 | 12.00 | 27 |
| 2005 | 4.6 | 14 | 12.00 | 27 |
| 2006 | 4.6 | 14 | 12.00 | 27 |
| 2007 | 4.7 | 14 | 16.80 | 60 |
| 2008 | 4.7 | 14 | 16.80 | 60 |
| 2009 | 4.7 | 14 | 16.80 | 60 |
| 2010 | 4.7 | 14 | 16.80 | 60 |
| 2011 | 4.7 | 14 | 16.80 | 60 |
| 2012 | 4.7 | 14 | 16.80 | 60 |
| 2013 | 4.7 | 14 | 16.80 | 60 |
| 2014 | 4.7 | 14 | 16.80 | 60 |
| 2015 | 4.7 | 14 | 16.80 | 60 |

**Table A.3: Postage Rates for first ounce of US First Class Mail**

| Date | Cost of First Ounce, Nominal USD<br><br>Source: (US Post Office Historian, 2018) | Cost of Additional Ounce(s), Nominal USD<br><br>Source: (Wikipedia, 2018f) | Cost of first ounce, (2016 USD) | Cost of additional ounce(s), (2016 USD) |
|---|---|---|---|---|
| November 1, 1981 | 20 | .17 | 0.51 | 0.44 |
| February 17, 1985 | 22 | .17 | 0.52 | 0.40 |
| April 3, 1988 | 25 | .2 | 0.62 | 0.43 |
| February 3, 1991 | 29 | .23 | 0.54 | 0.42 |
| January 1, 1995 | 32 | .23 | 0.52 | 0.37 |
| January 10, 1999 | 33 | .22 | 0.48 | 0.32 |
| January 7, 2001 | 34 | .21 | 0.47 | 0.29 |
| June 30, 2002 | 37 | .23 | 0.50 | 0.31 |
| January 8, 2006 | 39 | .24 | 0.48 | 0.29 |
| May 14, 2007 | 41 | .17 | 0.37 | 0.20 |
| May 12, 2008 | 42 | .17 | 0.48 | 0.20 |
| May 11, 2009 | 44 | .17 | 0.49 | 0.19 |
| January 22, 2012 | 45 | .20 | 0.47 | 0.21 |
| January 27, 2013 | 46 | .20 | 0.46 | 0.21 |
| January 26, 2014 | 49 | .21 | 0.50 | 0.21 |

**Table A.4  Total Annual Internet Usage**
From Sumits (2015)

| Year | Traffic (Gigabytes per year) |
|---|---|
| 1984 | 180 |
| 1985 | 396 |
| 1986 | 780 |
| 1987 | 1536 |
| 1988 | 3024 |
| 1989 | 5976 |
| 1990 | 12000 |
| 1991 | 24024 |
| 1992 | 53328 |
| 1993 | 104580 |
| 1994 | 309960 |
| 1995 | 1806000 |
| 1996 | 14400000 |
| 1997 | 60000000 |
| 1998 | 134400000 |
| 1999 | 306000000 |
| 2000 | 903000000 |
| 2001 | 2100000000 |
| 2002 | 4272000000 |
| 2003 | 8172600000 |
| 2004 | 15213600000 |
| 2005 | 21632947428 |
| 2006 | 34926952452 |
| 2007 | 53728412616 |
| 2008 | 77893913640 |
| 2009 | 1.11624E+11 |
| 2010 | 1.65012E+11 |
| 2011 | 2.39688E+11 |
| 2012 | 3.14579E+11 |
| 2013 | 3.93586E+11 |
| 2014 | 5.09078E+11 |

**Table A.5: Percentage of Internet use by Media Type**

| Year | Audio | Video |
|------|-------|-------|
| 1986 | 0.0%  | 0.0%  |
| 1987 | 0.0%  | 0.0%  |
| 1988 | 0.0%  | 0.0%  |
| 1989 | 0.0%  | 0.0%  |
| 1990 | 0.0%  | 0.0%  |
| 1991 | 0.0%  | 0.0%  |
| 1992 | 0.0%  | 0.0%  |
| 1993 | 4.5%  | 0.5%  |
| 1994 | 7.7%  | 1.3%  |
| 1995 | 9.9%  | 2.9%  |
| 1996 | 7.3%  | 6.6%  |
| 1997 | 7.4%  | 12.5% |
| 1998 | 9.5%  | 9.6%  |
| 1999 | 16.4% | 6.5%  |
| 2000 | 17.3% | 6.2%  |
| 2001 | 14.8% | 6.7%  |
| 2002 | 14.4% | 11.1% |
| 2003 | 11.6% | 13.8% |
| 2004 | 8.1%  | 20.5% |
| 2005 | 4.0%  | 27.2% |
| 2006 | 3.4%  | 32.7% |
| 2007 | 4.1%  | 36.6% |

Table A.5 is constructed by combining internet usage statistics for each type of media (Audio and Video) by combining the usage data sets from Hilman and Lopez (2011) and using equation B.1:

**Equation A.1**

$$\%IntUse_{Media}(t) = \sum_{Protocols} \%InternetUse_{Protocol\_i} * \%ProtocolUse_{Media}$$

Where Internet Protocols include:
- E-mail
- File Transfer Protocol (FTP)
- World Wide Web (www)
- Streaming, Peer-to-Peer (P2P)
- Direct Download

**Table A.6 Sales Numbers per Year for each of the Physical Media Types**
Derived from Hilbert and Lopez (2011).

| Year | Number of CDs (millions) | Number of Cassettes (millions) | Number of Vinyl Records (millions) | Number of DVDs (not including HD-DVD) (millions) | Number of VHS (+ VHS-C) (millions) |
|---|---|---|---|---|---|
| 1993 | 1183 | 625 | 80.4 | 0 | 662.5 |
| 1994 | 1789 | 610 | 56.3 | 0 | 782.7 |
| 1995 | 1885 | 500 | 38.1 | 0 | 902.9 |
| 1996 | 2071 | 450 | 31.8 | 0 | 1023.1 |
| 1997 | 2262 | 375 | 25.3 | 0.6 | 1043.5 |
| 1998 | 2384 | 350 | 22.3 | 2.7 | 905.6 |
| 1999 | 2499 | 275 | 21.2 | 10.3 | 819.4 |
| 2000 | 2556 | 200 | 19 | 17.9 | 649.5 |
| 2001 | 2441 | 150 | 14.6 | 23.3 | 479.7 |
| 2002 | 2324 | 100 | 9.8 | 26 | 455 |
| 2003 | 2213 | 50 | 8.2 | 109 | 423 |
| 2004 | 2186 | 25 | 7.8 | 325 | 347.3 |
| 2005 | 2055 | 10 | 5.9 | 943 | 276.1 |
| 2006 | 1981 | 5 | 5.6 | 1165 | 226.1 |
| 2007 | 1819 | 5 | 6.3 | 1431 | 150.3 |

**Table A.7: Number of Minutes per unit on Analog Physical Media**

| Media Type | Number of Minutes per Unit |
|---:|:---|
| VHS | 180 |
| Cassette | 60 |
| Vinyl | 90 |

**Table A.8: Amount of Data per Unit on Digital Physical Media**

| Media Type | Data per Unit (Megabytes) |
|---:|:---|
| CD | 700 |
| DVD | 4700 |